\documentclass[onecolumn,prl]{revtex4}
\usepackage{float}
\usepackage{makeidx}
\usepackage{graphicx,amsmath}

\begin{document}

\title{Higher Order Mode Entanglement in a Type II Optical Parametric Oscillator}

\author{Jun Guo, Chunxiao Cai, Long Ma, Kui Liu, Hengxin Sun, and Jiangrui Gao$^{\dagger}$}

\address{State Key Laboratory of Quantum Optics and Quantum Optics Devices, Institute of Opto-Electronics, Shanxi University, Taiyuan 030006, China}

\begin{abstract}

 Nonclassical beams in high order spatial modes have attracted much interest but they exhibit much less squeezing and entanglement than the fundamental spatial modes, limiting their applications. We experimentally demonstrate the relation between pump modes and entanglement of first-order HG modes (HG$_{10}$ entangled states) in a type II OPO and show that the maximum entanglement of high order spatial modes can be obtained by optimizing the pump spatial mode. To our knowledge, this is the first time to report this. Utilizing the optimal pump mode, the HG$_{10}$ mode threshold can be reached easily without HG$_{00}$ oscillation and HG$_{10}$ entanglement is enhanced by 53.5\% over HG$_{00}$ pumping. The technique is broadly applicable to entanglement generation in high order modes.

\end{abstract}

\maketitle

\section{Introduction}
Continuous variable (CV) squeezed and entangled states are important in processes such as quantum computation, quantum communication and quantum metrology. Since the 1985 observation of CV squeezing by Slusher et al. \cite{slusher85}, much research has followed on the generation and optimization of squeezing and entanglement in different systems. These include the optical parametric oscillator (OPO) \cite{wu86,Ou92}, four-wave mixing (FWM) \cite{slusher85}, and the in-fiber optical Kerr effect \cite{Drummond03,Friberg96}. Among these tools, the OPO is the most widely used. In recent years, squeezing of up to 15 dB in type I OPOs \cite{Vahlbruch16} and entanglement of 8.4 dB in type II OPOs \cite{Zhou15} were realized.

Traditionally most OPOs operate in the fundamental mode. However, higher order modes such as Hermite-Gauss (HG) and Laguerre-Gauss (LG) modes contain more spatial degrees of freedom and can give more information in applications than the fundamental mode. They can be used to enhance measurement precision of some physical quantities, such as lateral displacement \cite{Sun14} and transverse rotation angle of an optical beam \cite{Ambrosio13}. They can also be applied in quantum imaging \cite{Brida10}, quantum storage \cite{Nicolas14}, quantum super-dense coding \cite{Barreiro08}, and biological measurement \cite{Taylor13}. In recent years, squeezing and entanglement have been expanded to higher order modes in OPOs. Lassen et al. generated quadrature squeezing of HG$_{00}$, HG$_{10}$ and HG$_{20}$ modes separately with a type I OPO in 2006 \cite{Lassen06,Lassen07} and quadrature entanglement of first-order LG modes with a type I OPO in 2009 \cite{Lassen09}. Multimode squeezing and entanglement can also be generated in a specially designed OPO \cite{Janousek09,Chalopin10,Chalopin11}. Recently, a CV hyperentanglement state, wherein both spin and orbital angular momenta are entangled, was realized in a multimode type II OPO \cite{Santos09,Liu14}.

To date the degree of squeezing and entanglement produced in higher order modes has been much lower than for the fundamental mode, which limits their applications. Almost all the above cited work adopted the fundamental mode as the pump for the higher order signal modes. This lead to low pump conversion efficiencies and crucially much higher oscillation thresholds than for the fundamental spatial mode, severely limiting the attainable squeezing and entanglement levels.

Lassen et al. presented the ideal pump for oscillation of the HG$_{10}$ mode, a superposition of HG$_{00}$ and HG$_{20}$ modes, but synthesizing the multi-mode is experimentally very challenging \cite{Lassen06,Lassen07,Lassen09}. In this Express paper, we experimentally demonstrate the relation between pump modes and entanglement of first-order HG modes (HG$_{10}$ entangled states) in a type II OPO and show that the maximum entanglement of high order spatial modes can be obtained by optimizing the pump spatial mode. To our knowledge, this is the first time to report this. Using the optimal pump, the entanglement inseparability for HG$_{10}$ mode is enhanced by 53.5\% and the threshold is reduced by 66.7\% relative to using HG$_{00}$ in our result.

\section{Theoretical model}
For a type II OPO with an HG$_{10}$ signal mode, we define $v^{p}(\vec{r})$  as the transverse distribution of the pump mode where $\vec{r}=(x,y)$ denotes the transverse coordinates. This can be expanded into a series of HG modes as
\begin{equation}
v^{p}(\vec{r})=\sum\limits _{n=0}^{\infty}{c_{n}v_{n0}(\vec{r})},
\end{equation}
where $v_{n0}(\vec{r})$ denotes the transverse profile of the nth order HG mode and $c_{n}$ is its corresponding coefficient. The transverse profiles of the signal and idler modes can be described by $ u^s \left( {\vec r} \right) $ and $ u^i \left( {\vec r} \right) $. The full Hamiltonian of the system can be written as
\begin{equation}
\hat{H}=i\hbar\varepsilon^{p}\left({\hat{a}^{p\dag}-\hat{a}^{p}}\right)+i\hbar\chi\Gamma\left({\hat{a}^{p}{\text{ }}\hat{a}^{s\dag}{\text{ }}\hat{a}^{i\dag}-\hat{a}^{p\dag}\hat{a}^{s}{\text{ }}\hat{a}^{i}}\right),
\end{equation}
where $\chi$ is the nonlinear coefficient of the crystal, $\hat{a}^{p}$, $\hat{a}^{s}$ and $\hat{a}^{i}$ are the annihilation operators of the pump, signal and idler fields, and $\varepsilon^{p}$ is the pump parameter. $\Gamma$ is the coupling coefficient of the three intracavity fields given by
\begin{equation}
\Gamma=\int_{-\infty}^{+\infty}{v^{p}\left({\vec{r}}\right)u^{s*}\left({\vec{r}}\right)u^{i*}
\left({\vec{r}}\right)d\vec{r}}.
\end{equation}

Additionally considering the quantum vacuum noise caused by the extra losses, the Langevin equations of motion for the intracavity fields can be given by
\begin{subequations}
\begin{align}
\tau \dot \hat a^p \left( t \right) = - \gamma _p \hat a^p \left( t \right) - \chi \Gamma \hat a^s \left( t \right)\hat a^i \left( t \right) + \varepsilon ^p e^{ - i\theta _p } + \sqrt {2\mu _p } \hat b_{in}^p \left( t \right),  \\
\tau \dot \hat a^s \left( t \right) = - \gamma '_s \hat a^s \left( t \right) + \chi \Gamma \hat a^p \left( t \right)\hat a^{i\dag } \left( t \right) + \sqrt {2\gamma _s } \hat a^s _{in} \left( t \right) + \sqrt {2\mu _s } \hat b^s _{in} \left( t \right),\\
\tau \dot \hat a^i \left( t \right) = - \gamma '_i \hat a^i \left( t \right) + \chi \Gamma \hat a^p \left( t \right)\hat a^{s\dag } \left( t \right) + \sqrt {2\gamma _i } \hat a^i _{in} \left( t \right) + \sqrt {2\mu _i } \hat b^i _{in} \left( t \right)  .
\end{align}
\end{subequations}

Here $\gamma_{k}$ ($k=p, s, i$) are the transmission losses through the output coupler and $\mu_{k}$ are all other extra losses, $ \gamma '_k = \gamma _k + \mu _k \left( {k = s,i} \right) $ are the total losses. $ \tau $ is the round-trip time of the three modes in the cavity, $\theta_{p}$ is the phase of the pump field, $ \hat a^l _{in} \left( t \right) $ ($l=s, i$) are the input signal and idler fields, and $ \hat b_{in}^m \left( t \right) $ ($m=p,s,i$) are the quantum vacuum noise of the three fields induced by the extra losses. Assuming the loss factors $\gamma_{p}=1$, $\gamma_{s}=\gamma_{i}=\gamma$, $\mu_{s}=\mu_{i}=\mu$ and $\gamma'_{s}=\gamma'_{i}=\gamma'$, then the oscillation threshold is obtained as
\begin{equation}
\varepsilon^{pth}={{\gamma'}\mathord{\left/{\vphantom{{\gamma'}{\left({\chi\Gamma}\right)}}}\right.\kern -\nulldelimiterspace}{\left({\chi\Gamma}\right)}}.
\end{equation}

$\left\langle {\hat a_{in}^l } \right\rangle = \alpha _{in}^l e^{ - i\theta _l } \left( {l = s,i} \right) $, where $\theta_{l}$  are the phases of the input signal and idler fields. We introduce the amplitude quadrature $\hat{X}={{\left({\hat{a}+\hat{a}^{\dag}}\right)}\mathord{\left/{\vphantom{{\left({\hat{a}+\hat{a}^{\dag}}\right)}2}}\right.\kern -\nulldelimiterspace}2}$ and phase quadrature $\hat{Y}={{-i\left({\hat{a}-\hat{a}^{\dag}}\right)}\mathord{\left/{\vphantom{{-i\left({\hat{a}-\hat{a}^{\dag}}\right)}2}}\right.\kern -\nulldelimiterspace}2}$. When the relative phase between the pump and the seed $\varphi=\theta_{p}-\left({\theta_{s}+\theta_{i}}\right)=0$, the system is in a parametric amplification state, and the correlation noise spectra can be given by
\begin{equation}
V_{\hat{X}^{s}-\hat{X}^{i}}=V_{\hat{Y}^{s}+\hat{Y}^{i}}=1-\eta_{esc}\frac{{4\sigma}}{{\left({1+\sigma}\right)^{2}+\Omega^{2}}},
\end{equation}
where $\eta_{esc}=\gamma/\gamma'$ is the escape efficiency, $\sigma=\varepsilon^{p}/\varepsilon^{pth}$ is the normalized pump parameter, and $\Omega=\omega\tau/\gamma'$ is the normalized analyzing frequency. When the relative phase between the pump and the seed $\varphi=\theta_{p}-\left({\theta_{s}+\theta_{i}}\right)=\pi$, the system is in a parametric deamplification state, and the correlation noise spectra can be given by
\begin{equation}
V_{\hat{X}^{s}+\hat{X}^{i}}=V_{\hat{Y}^{s}-\hat{Y}^{i}}=1-\eta_{esc}\frac{{4\sigma}}{{\left({1+\sigma}\right)^{2}+\Omega^{2}}},
\end{equation}

Considering the total detection efficiency of the system, $\eta_{det}$, Eq. (7) can be rewritten as
\begin{equation}
V_{\hat{X}^{s}+\hat{X}^{i}}=V_{\hat{Y}^{s}-\hat{Y}^{i}}=1-\eta_{\det}\eta_{esc}\frac{4\sqrt{p/p_{th}}}{\left(1+\sqrt{p/p_{th}}\right)^{2}+\Omega^{2}},
\end{equation}
where $\eta_{det}=\eta_{prop}\eta_{hd}\eta_{phot}$, $\eta_{prop}$ is the propagation efficiency, $\eta_{hd}$ is the homodyne detection efficiency and $\eta_{phot}$ is the quantum efficiency of the photodiode. The normalized pump power is given by $p/p_{th}=\sigma^{2}$, where $p$ is the actual pump power and $p_{th}={{\gamma'^{2}}\mathord{\left/{\vphantom{{\gamma'^{2}}{\left({\chi^{2}\Gamma^{2}}\right)}}}\right.\kern -\nulldelimiterspace}{\left({\chi^{2}\Gamma^{2}}\right)}}$ is the threshold pump power.

The inseparability criterion can be expressed as \cite{Duan00}
\begin{equation}
V=V_{X_{}^{s}+X_{}^{i}}+V_{Y_{}^{s}-Y_{}^{i}}=2-\eta_{\det}\eta_{esc}\frac{{8\sqrt{{p\mathord{\left/{\vphantom{p{p_{th}}}}\right.\kern -\nulldelimiterspace}{p_{th}}}}}}{{\left({1+\sqrt{{p\mathord{\left/{\vphantom{p{p_{th}}}}\right.\kern -\nulldelimiterspace}{p_{th}}}}}\right)^{2}+\Omega^{2}}}<2.
\end{equation}

From Eq. (3), (4) and (5), different pump modes correspond to different coupling coefficients and thus different nonlinear efficiencies, leading to different pump thresholds. The coupling coefficient for the HG$_{00}$ signal mode $u_{00}\left({\vec{r}}\right)$ with HG$_{00}$ pump mode is
\begin{equation}
\Gamma=\int_{-\infty}^{+\infty}{v_{00}\left({\vec{r}}\right)\left[{u_{00}\left({\vec{r}}\right)}\right]^{2}d\vec{r}}=1,
\end{equation}
so the oscillation threshold for the HG$_{00}$ signal mode with HG$_{00}$ pump is $p_{th}^{00\to00}={{\gamma'^{2}}\mathord{\left/{\vphantom{{\gamma'^{2}}{\left({\chi^{2}\Gamma^{2}}\right)}}}\right.\kern -\nulldelimiterspace}{\left({\chi^{2}\Gamma_{}^{2}}\right)}}={{\gamma'^{2}}\mathord{\left/{\vphantom{{\gamma'^{2}}{\chi^{2}}}}\right.\kern -\nulldelimiterspace}{\chi^{2}}}$. For the HG$_{10}$ signal mode $u_{10}\left({\vec{r}}\right)$ generation with all possible pump, we have the expression from Eq. (1)
\begin{equation}
\Gamma=\sum\limits _{n=0}^{\infty}{c_{n}}\int_{-\infty}^{+\infty}{v_{n0}\left({\vec{r}}\right)\left[{u_{10}^{}\left({\vec{r}}\right)}\right]^{2}d\vec{r}}=\sum\limits _{n=0}^{\infty}{c_{n}\Gamma_{n}},
\end{equation}
where $\Gamma_{n}=\int_{-\infty}^{+\infty}{v_{n0}\left({\vec{r}}\right)\left[{u_{10}\left({\vec{r}}\right)}\right]^{2}d\vec{r}} $ denotes the coupling coefficient of the nth order HG pump mode. These are
\begin{equation}
\Gamma_{0}=\int_{-\infty}^{+\infty}{v_{00}\left({\vec{r}}\right)\left[{u_{10}\left({\vec{r}}\right)}\right]^{2}d\vec{r}}={1\mathord{\left/{\vphantom{12}}\right.\kern -\nulldelimiterspace}2},
\end{equation}

\begin{equation}
\Gamma_{2}=\int_{-\infty}^{+\infty}{v_{20}\left({\vec{r}}\right)\left[{u_{10}\left({\vec{r}}\right)}\right]^{2}d\vec{r}}={1\mathord{\left/{\vphantom{1{\sqrt{2}}}}\right.\kern -\nulldelimiterspace}{\sqrt{2}}},
\end{equation}
and $\Gamma_{n}=0$ for all other $n$. The HG$_{10}$ signal mode threshold with an HG$_{00}$ pump mode $\left({c_{0}=1}\right)$ is $p_{th}^{00\to10}={{\gamma'^{2}}\mathord{\left/{\vphantom{{\gamma'^{2}}{\left({\chi^{2}\Gamma_{0}^{2}}\right)}}}\right.\kern -\nulldelimiterspace}{\left({\chi^{2}\Gamma_{0}^{2}}\right)}}{{=4\gamma'^{2}}\mathord{\left/{\vphantom{{=4\gamma'^{2}}{\chi^{2}}}}\right.\kern -\nulldelimiterspace}{\chi^{2}}}$, and with an HG$_{20}$ pump mode $\left({c_{2}=1}\right)$ it is $ p_{th}^{20\to10}={{{{\gamma'^{2}}\mathord{\left/{\vphantom{{\gamma'^{2}}{\left({\chi^{2}\Gamma_{2}^{2}}\right)}}}\right.\kern -\nulldelimiterspace}{\left({\chi^{2}\Gamma_{2}^{2}}\right)}}=2\gamma'^{2}}\mathord{\left/{\vphantom{{{{\gamma'^{2}}\mathord{\left/{\vphantom{{\gamma'^{2}}{\left({\chi^{2}\Gamma_{2}^{2}}\right)}}}\right.\kern -\nulldelimiterspace}{\left({\chi^{2}\Gamma_{2}^{2}}\right)}}=2\gamma'^{2}}{\chi^{2}}}}\right.\kern -\nulldelimiterspace}{\chi^{2}}}$.

For the optimal pump mode, $\Gamma=c_{0}\Gamma_{0}+c_{2}\Gamma_{2}=\left({1\mathord{\left/{\vphantom{12}}\right.\kern -\nulldelimiterspace}2}\right)c_{0}+\left({1\mathord{\left/{\vphantom{1{\sqrt{2}}}}\right.\kern -\nulldelimiterspace}{\sqrt{2}}}\right)c_{2}$, and $c_{0}^{2}+c_{2}^{2}=1$. The maximum value of $\Gamma$ is ${{\sqrt{3}}\mathord{\left/{\vphantom{{\sqrt{3}}2}}\right.\kern -\nulldelimiterspace}2}$, with $c_{0}=\sqrt{{1\mathord{\left/{\vphantom{13}}\right.\kern -\nulldelimiterspace}3}}$ and $c_{2}=\sqrt{{2\mathord{\left/{\vphantom{23}}\right.\kern -\nulldelimiterspace}3}}$, so the optimal pump mode is $v^{p}=\sqrt{{1\mathord{\left/{\vphantom{13}}\right.\kern -\nulldelimiterspace}3}}v_{00}+\sqrt{{2\mathord{\left/{\vphantom{23}}\right.\kern -\nulldelimiterspace}3}}v_{20}$, a superposition of HG$_{00}$ and HG$_{20}$ modes. The HG$_{10}$ signal mode threshold with the optimal pump mode is ${{p_{th}^{opt\to10}=\gamma'^{2}}\mathord{\left/{\vphantom{{p_{th}^{opt\to10}=\gamma'^{2}}{\left({\chi^{2}\Gamma_{}^{2}}\right)}}}\right.\kern -\nulldelimiterspace}{\left({\chi^{2}\Gamma_{}^{2}}\right)}}{{=4\gamma'^{2}}\mathord{\left/{\vphantom{{=4\gamma'^{2}}{3\chi^{2}}}}\right.\kern -\nulldelimiterspace}{3\chi^{2}}}$.

\begin{figure}[ht!]
\centering\includegraphics[width=10cm]{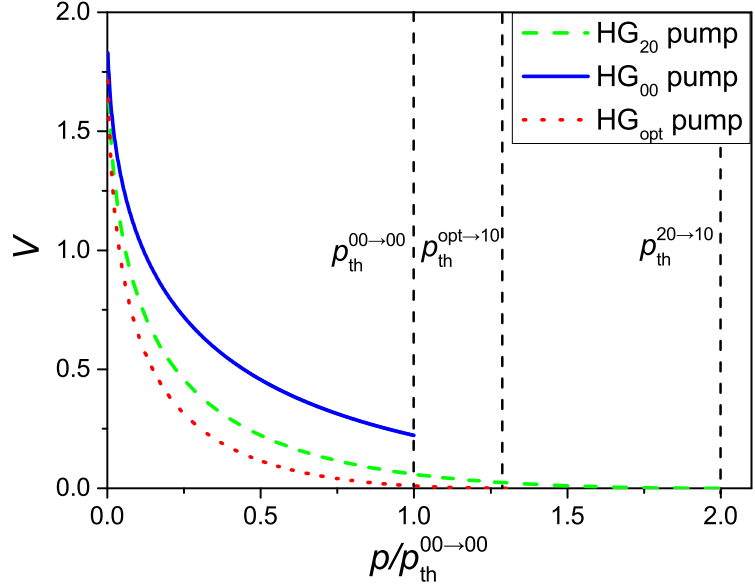}
\caption{Theoretical inseparabilities \emph{V} against normalized pump power $ p/p_{th}^{00 \to 00} $ for three pump modes, HG$_{00}$ (blue solid line), HG$_{20}$ (green dashed line) and the optimal pump mode HG$_\mathrm{opt}$ (red dotted line) under ideal conditions. The parameters are $\eta_{\det}=1$, $\eta_{esc}=1$, $\Omega=0$.}
\end{figure}

Fig. 1 gives the theoretical curves of the inseparabilities versus normalized pump power for the three different pump modes HG$_{00}$, HG$_{20}$, and the optimal superposition under ideal conditions. Under HG$_{00}$ pumping, the HG$_{00}$ signal mode threshold is $p_{th}^{00\to00}$, which is one-quarter that of the HG$_{10}$ signal mode $p_{th}^{00\to10}=4p_{th}^{00\to00}$. When the pump power reaches the HG$_{00}$ threshold $p_{th}^{00\to00}$, the system starts to oscillate in the HG$_{00}$ mode, so the maximum HG$_{10}$ entanglement cannot be obtained. However, with HG$_{20}$ pumping, the HG$_{00}$ signal mode will not be excited. The HG$_{10}$ signal mode threshold $p_{th}^{20\to10}=2p_{th}^{00\to00}$  can be reached with enough pump power in theory, so the maximum HG$_{10}$ entanglement can be obtained using an HG$_{20}$ pump. With optimal superposition mode pumping, the HG$_{00}$ pump mode comprises 1/3 the total pump power. The threshold for the HG$_{10}$ signal mode is $p_{th}^{opt\to10}={{4p_{th}^{00\to00}}\mathord{\left/{\vphantom{{4p_{th}^{00\to00}}3}}\right.\kern -\nulldelimiterspace}3}$. Hence the maximum power of the HG$_{00}$ component of the pump is ${{4p_{th}^{00\to00}}\mathord{\left/{\vphantom{{4p_{th}^{00\to00}}9}}\right.\kern -\nulldelimiterspace}9}$, which is much smaller than the HG$_{00}$ signal mode threshold $p_{th}^{00\to00}$. The HG$_{00}$ signal mode will therefore not oscillate in under optimal mode pumping. Moreover, since the HG$_{10}$ signal mode threshold is much lower than for pure HG$_{20}$ pumping, the maximum entanglement can be obtained at lower pump power.

\section{Experiment}
\begin{figure}[ht!]
\centering\includegraphics[width=10cm]{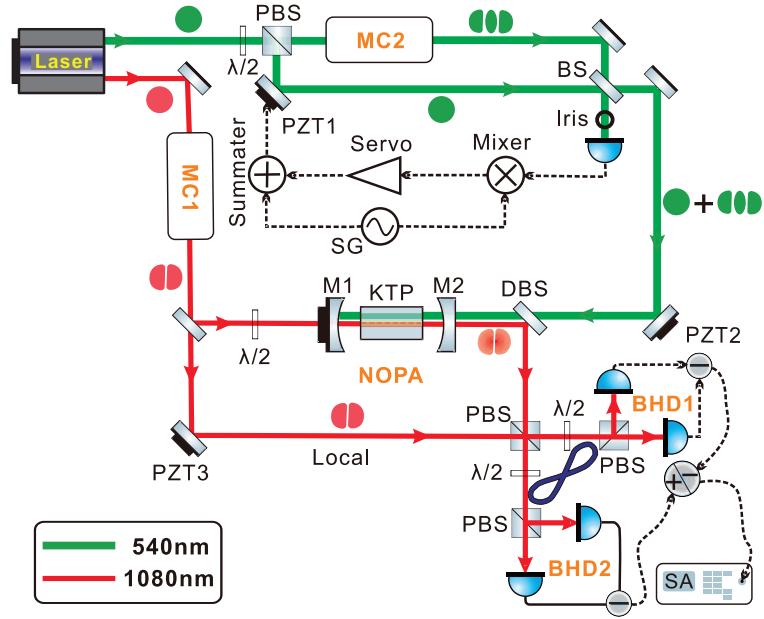}
\caption{Schematic of the experimental setup. NOPA: non-degenerate optical parametric amplifier, KTP: type II KTP crystal, M1 and M2: cavity mirrors, BS: beam splitter, PBS: polarizing beam splitter, MCs: mode converters, SG: signal generator, Servo: servo amplifier circuit for feedback system, PZTs: piezoelectric transducers, DBS: dichroic beam splitter, Local: local oscillator, BHDs: balanced homodyne detectors, +/-: positive/negative power combiner, and SA: spectrum analyzer.}
\end{figure}
The experimental setup is depicted in Fig. 2. A continuous wave all solid state laser source emits both infrared at 1080 nm and green light at 540 nm. The infrared beam passes through a mode converter (MC1), which converts the HG$_{00}$ mode into the HG$_{10}$ mode. A part of the HG$_{10}$ mode is injected into a non-degenerate optical parametric amplifier (NOPA) as the seed beam, and the rest of it is used as the local oscillator for homodyne detection. The green beam is used as the pump beam. It is split into two, one beam pass through the mode converter MC2, which converts HG$_{00}$ mode into HG$_{20}$ mode, the other beam is still HG$_{00}$ mode, then the two beams are combined by a beamsplitter, generating the superposition pump mode. By this arrangement, we can choose to pass either the HG$_{00}$, the HG$_{20}$, or the superposition pump mode.

To lock the relative phase between the HG$_{00}$ and HG$_{20}$ modes, we use an iris aperture to pass only the center of the beam profile to a photodiode. With a lock-in amplifier, the relative phase is locked to zero. The mode converters and the NOPA cavity are locked using the standard Pound-Drever-Hall (PDH) technique \cite{Drever83}.

The NOPA cavity consists of two 30 mm radius of curvature plano-concave mirrors and a $3\times3\times10$ $mm^{3}$ type II KTP crystal in the center. The seed beam input mirror M1 is highly reflective (R>99.95\%) at both 1080 nm and 540 nm. The transmittance T of the output coupler M2 is 6\% at 1080 nm and T>95\% at 540 nm. The cavity is nearly concentric with a length of 62.5 mm and has a waist of 41 $\mu m$ in the infrared and 29 $\mu m$ in the green. The NOPA has a finesse of 84 for the signal beam with a free spectral range of 2.4 GHz and a bandwidth of 28 MHz. We lock the relative phase between the seed and the pump beam in the parametric deamplification regime with PZT2.

The NOPA output beams and the green beam pass through a dichroic beam splitter (DBS), which reflects only the infrared beam to be measured. This is divided into two parts by a PBS. They are detected by two balanced homodyne detectors (BHDs). The photocurrents from the two BHDs feed a positive/negative combiner (+/-), and those outputs are recorded by a spectrum analyzer (SA). The correlation noise spectra of the amplitude sum and phase difference of the signal and idler beams are measured by scanning the phase of the local infrared beam using a mirror mounted on piezoelectric transducer PZT3.

\section{Experimental results}
The experimental parameters in our experiment are as follows. The analyzing frequency is 5 MHz, the resolution bandwidth (RBW) is 300 kHz, and the video bandwidth (VBW) is 1 kHz. The bandwidth of the NOPA is 28 MHz (from which $\Omega$  = 5 MHz/28 MHz = 0.18). The various efficiencies are $\eta_{prop}$ = 0.89$\pm$0.02, $\eta_{phot} $ = 0.90$\pm$0.01, $\eta_{hd}$ = 0.81$\pm$0.02, and $\eta_{esc}$ = 0.79$\pm$0.01, thus the total efficiency $\eta_{total}$  = 0.51$\pm$0.04. The pump threshold for the HG$_{00}$ signal mode with an HG$_{00}$ pump is $p_{th}^{00\to00}$ = 510 mW. From theoretical prediction, the oscillation threshold for the HG$_{10}$ signal mode is $p_{th}^{00\to10}$ = 2.04 W with HG$_{00}$ pumping, it is $p_{th}^{20\to10}$ = 1.02 W with HG$_{20}$ pumping, and with the optimal superposition mode pumping $p_{th}^{opt\to10}$  = 680 mW.

The measured entanglement inseparabilities \emph{V} are plotted against the normalized pump power $p/p_{th}^{00\to00}$ for the three different pump modes in Fig. 3. The corresponding theoretical curves in experimental conditions are also depicted.

\begin{figure}[ht!]
\centering\includegraphics[width=13cm]{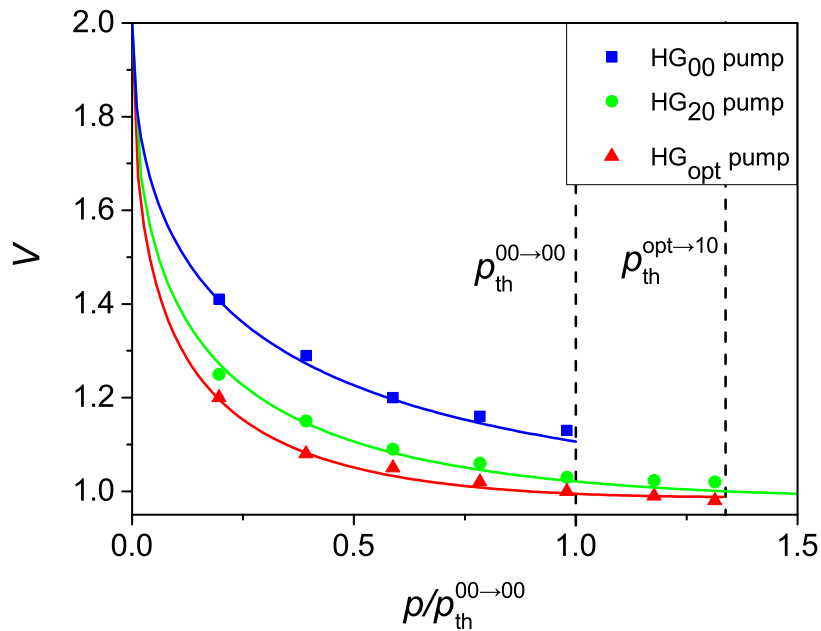}
\caption{The inseparabilities \emph{V} versus normalized pump power $p/p_{th}^{00\to00}$, where $p_{th}^{00\to00}$ = 510 mW. Data points from the experiment are blue squares for HG$_{00}$ pumping, green circles for HG$_{20}$ pumping and red triangles for the optimal pump mode HG$_\mathrm{opt}$. The solid curves are the theoretical values in experimental conditions for the three pump modes.}
\end{figure}

From Fig. 3, the entanglement increases with the increasing pump power for the three pump modes and there is good agreement between theory and experiment. At a given pump power, the optimal pump mode HG$_\mathrm{opt}$ outperforms the other two modes and the HG$_{20}$ pump mode outperforms HG$_{00}$. However, the minimum value of \emph{V} is not close to zero as fig.1 due to the nonideal cavity and detection system. The maximum pump power for  HG$_{00}$ mode in our experiment is 500 mW, since the oscillating threshold of the HG$_{00}$ signal mode is 510 mW, at higher power, the OPO will oscillate on the HG$_{00}$ mode, so the maximum entanglement of HG$_{10}$ mode can not be obtained with HG$_{00}$ pump mode. However, with HG$_{20}$ pump mode or the optimal pump mode, the maximum entanglement of HG$_{10}$ mode can be obtained. Moreover, with the optimal pump mode HG$_\mathrm{opt}$, the maximum entanglement can be obtained at lower pump power compared with HG$_{20}$ pumping.

\begin{figure}[ht!]
\centering\includegraphics[width=13cm]{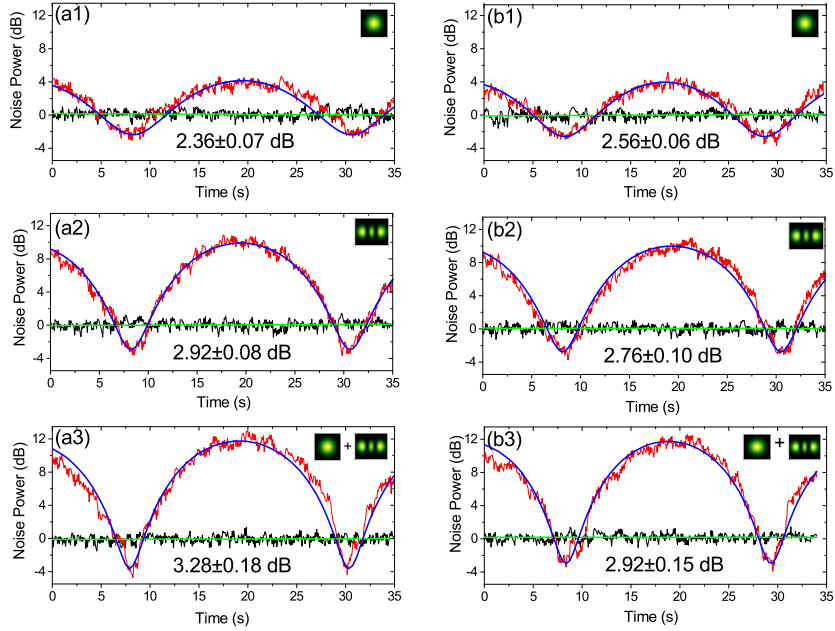}
\caption{The HG$_{10}$ mode correlation noise powers for the amplitude sum (a1-a3) and the phase difference (b1-b3). The top row was taken using 500 mW of HG$_{00}$ pumping, the middle row with 670 mW of HG$_{20}$ pumping, and the bottom row with 670 mW of the superposition HG$_\mathrm{opt}$ pumping.}
\end{figure}

Fig. 4 gives the HG$_{10}$ mode correlation noise powers with the three different pump modes. For HG$_{00}$ pumping at 500 mW, the amplitude sum power was 2.36$\pm$0.07 dB and the phase difference power was 2.56$\pm$0.06 dB. For HG$_{20}$ pumping at 670 mW these were 2.92$\pm$0.08 dB and 2.76$\pm$0.10 dB. For HG$_\mathrm{opt}$ pumping at 670 mW, the powers were 3.28$\pm$0.18 dB and 2.92$\pm$0.15 dB. The HG$_{10}$ mode entanglement inseparabilities for the three pump modes are
\begin{equation}
V_{00}=\left\langle {\Delta^{2}\left({\hat{X}_{10}^{s}+\hat{X}_{10}^{i}}\right)}\right\rangle +\left\langle {\Delta^{2}\left({\hat{Y}_{10}^{s}-\hat{Y}_{10}^{i}}\right)}\right\rangle =1.13\pm0.02<2
\end{equation}
\begin{equation}
V_{20}=\left\langle {\Delta^{2}\left({\hat{X}_{10}^{s}+\hat{X}_{10}^{i}}\right)}\right\rangle +\left\langle {\Delta^{2}\left({\hat{Y}_{10}^{s}-\hat{Y}_{10}^{i}}\right)}\right\rangle =1.04\pm0.02<2
\end{equation}
\begin{equation}
V_{opt}=\left\langle {\Delta^{2}\left({\hat{X}_{10}^{s}+\hat{X}_{10}^{i}}\right)}\right\rangle +\left\langle {\Delta^{2}\left({\hat{Y}_{10}^{s}-\hat{Y}_{10}^{i}}\right)}\right\rangle =0.98\pm0.04<2
\end{equation}
Considering the total detection efficiency $\eta_{\det}=\eta_{prop}\eta_{phot}\eta_{hd}=0.65\pm0.04$, the inseparabilities of Eqs. (14-16) become 0.66$\pm$0.03, 0.52$\pm$0.03 and 0.43$\pm$0.06. Compared with HG$_{00}$ pumping, the inseperability is enhanced by $\eta=53.5\%$ using the optimal pump mode.

Summarizing the experimental results, we cannot obtain the maximum entanglement of the HG$_{10}$ mode with HG$_{00}$ pumping because of the low HG$_{00}$ threshold. With HG$_{20}$ pumping, this is not the case. Theoretically, the HG$_{10}$ signal mode threshold can be reached and the maximum entanglement can be obtained, but in our experiment the laser-limited pump power is insufficient. With the optimal pump mode, the HG$_{10}$ signal mode threshold is lower and the maximum entanglement can be obtained with lower pump power. Experimentally however, generating the optimal pump mode is relatively complicated and somewhat difficult. Using HG$_{20}$ pumping is operationally much easier and with sufficient power we can obtain the same degree of entanglement as the optimal pump mode.

\section{Conclusion}
We experimentally studied HG$_{10}$ mode entanglement in a type II OPO with three pump modes, HG$_{00}$, HG$_{20}$, and a superposition of the two modes. The superposition mode, a one-third HG$_{00}$ and two-thirds HG$_{20}$ combination, is theoretically optimal and experimentally shown to be able to obtain a higher entanglement at lower pump power. The experimental results match the theoretical prediction very well. The degree of entanglement is still relatively low resulting from extra losses and various inefficiencies in our experiment. The technique holds promise to obtain more than 10 dB squeezing for applications in quantum imaging \cite{Taylor16,Tsang16}. It is an efficient way to improve the squeezing of high-order spatial modes. Moreover, the method can be extended to high-dimension orbital angular momentum entanglement \cite{Liu16,Pan12,Zhou16} to enhance the generation efficiency.

\section*{Funding}
Ministry of Science and Technology of the People's Republic of China (MOST) (2016YFA0301404); National Natural Science Foundation of China (NSFC) (91536222, 61405108,11674205 ); NSFC Project for Excellent Research Team (61121064); University Science and Technology Innovation Project in Shanxi Province (2015103).

\end{document}